%% file: acl_latex.tex
\pdfoutput=1

\documentclass[11pt]{article}

\usepackage[]{acl}

\usepackage{times}
\usepackage{latexsym}
\usepackage{amsmath}
\usepackage{booktabs}
\usepackage[T1]{fontenc}
\usepackage{tabularx}

\usepackage[utf8]{inputenc}
\usepackage{graphicx}
\usepackage{subfig}

\usepackage{microtype}

\usepackage{inconsolata}

\title{Selecting Query-bag as Pseudo Relevance Feedback for \\ Information-seeking Conversations}

\author{
\begin{tabular}{c}
Xiaoqing Zhang$^{1}$ \quad \quad  Xiuying Chen$^{2}$ \\\textbf{Shen Gao}$^{3}$ \quad \textbf{Shuqi Li}$^{1}$ \quad \textbf{Xin Gao}$^{2}$\\ 
\textbf{Ji-Rong Wen}$^{1}$ \ \quad \quad \ \ \textbf{Rui \ Yan}$^{1}$\thanks{\ \ Corresponding authors.} 
\end{tabular}
\\ \vspace{.5mm}
    \small
    \begin{tabular}{c}
    $^1$Gaoling School of Artificial Intelligence, Renmin University of China \\
    $^2$King Abdullah University of Science and Technology\\
    $^3$Shandong University\\
    \end{tabular}
    \\ \vspace{.5mm}
    \small
    \begin{tabular}{c}
    \texttt{\{xiaoqingz, shuqili, jrwen, ruiyan\}@ruc.edu.cn} \\
    \texttt{\{xiuying.chen, xin.gao\}@kaust.edu.sa}\\
    \texttt{\{shengao\}@sdu.edu.cn}\\
    \end{tabular}
    \vspace{2mm} \\
}

\usepackage{multirow}
\usepackage{siunitx}
\usepackage{booktabs}
\begin{document}
\maketitle
\begin{abstract}
Information-seeking dialogue systems are widely used in e-commerce systems, with answers that must be tailored to fit the specific settings of the online system.
Given the user query, the information-seeking dialogue systems first retrieve a subset of response candidates, then further select the best response from the candidate set through re-ranking.
Current methods mainly retrieve response candidates based solely on the current query, however, incorporating similar questions could introduce more diverse content, potentially refining the representation and improving the matching process.
Hence, in this paper, we proposed a Query-bag based Pseudo Relevance Feedback framework (QB-PRF), which constructs a query-bag with related queries to serve as pseudo signals to guide information-seeking conversations.
Concretely, we first propose a Query-bag Selection module (QBS), which utilizes contrastive learning to train the selection of synonymous queries in an unsupervised manner by leveraging the representations learned from pre-trained VAE.
Secondly, we come up with a Query-bag Fusion module (QBF) that fuses synonymous queries to enhance the semantic representation of the original query through multidimensional attention computation.
We verify the effectiveness of the QB-PRF framework on two competitive pretrained backbone models, including BERT and GPT-2. 
Experimental results on two benchmark datasets show that our framework achieves superior performance over strong baselines\footnote{Code and checkpoints will be released upon acceptance.}.
\end{abstract}

\begin{figure}[ht]
    \centering
    \includegraphics[width=1\linewidth]
    {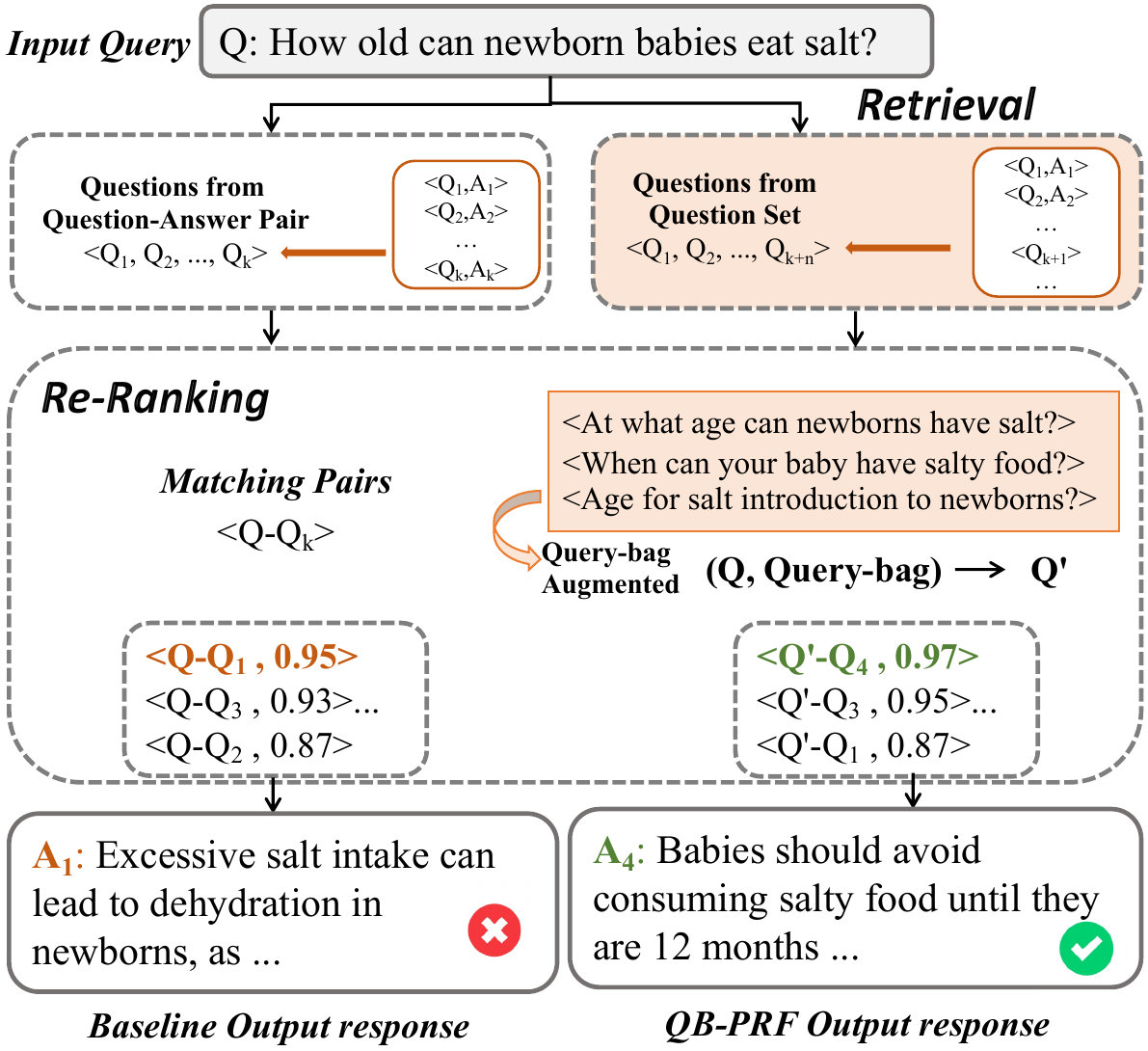}
    \caption{The comparison of existing baseline and our framework for information-seeking conversation.
    We use query-bag to enhance the query representation so as to enhance re-ranking performance.}
    \label{fig:intro}
    \vspace{-3mm}
\end{figure}

\section{Introduction}
Human-machine conversational systems in e-commerce differ significantly from open-domain dialogue systems. 
The responses they provide are tailored specifically to the context of the online shopping platform in question. 
For instance, the precise process of executing a particular action can vary between different platforms such as Amazon and eBay. 
Hence, there has been considerable research dedicated to retrieval-based systems in this area~\cite{qiu2018transfer,solomon1997conversation,fu2020context}, to retrieve an accurate answer from a curated answer pool.
Concretely, given a query, these systems 1) retrieve a subset of response candidates through pre-defined relevance searching processes between the query’s context and the response’s context, then further 2) select the best response from the candidate set through re-ranking methods, that are normally accomplished by computing the matching degree between the query’s context and the response \cite{fu2020context,wu2016sequential,ma2019triplenet,yan2016learning,zhou2018multi}.

Most existing models rely solely on the current query for retrieval and selection, which can be limiting given the diversity of expression for any single meaning. 
For example, the queries ``How old can newborn babies eat salt?'' and ``Suitable salt intake for newborns?'' convey similar intents but use different wording. 
Frequently, numerous related queries share similar meanings and can be utilized to enhance the representation of the query, thereby improving the current question answering system's comprehension and responses. The Figure \ref{fig:intro} illustrates a comparison between information-seeking conversations with and without query-bag.
However, it is not easy to construct such ``query bag'' with similar meanings.
The first challenge involves retrieving a diverse range of synonymous response candidates, given that the entire dataset may be noisy and include queries that are similar but not synonymous.
Once we have a diversified set of queries, the second challenge is to effectively combine this query collection with the original query to create an enhanced representation.

To resolve the above challenges, in this paper, we proposed a training paradigm called Query-bag  Pseudo Relevance Feedback (QB-PRF) for Information-seeking Conversations, which can be applied in any downstream matching models.
QB-PRF includes a Query-bag Selection Module (QBS) for choosing synonymous queries and a Query-bag Fusion Module (QBF) to enhance the query representation.
Specifically, we employs a contrastive mechanism to develop a distinct semantic representation of queries. 
The representation is then used in QBS to identify and select similar queries, i.e., a query-bag, for information enhancement, thereby improving the accuracy of retrieval.
With the query-bag, we can learn from distinctive query representation, which could serve as a pseudo supervision signal and ensure the framework selects valuable and diversified data with the same meaning to form the query-bag.

For QBF, it learns to fuse the query-bag information received by the QBS module to improve the performance of text-matching tasks.
Specifically, we applied the powerful transformer model to learn segment disambiguation and query-bag fusion. 
By doing so, we could enrich the semantic representation and learn a more generalized representation of the queries.
The fused representation will then serve as the new query representation and will be further passed to any matching model for boosting the performance of text-matching tasks.
The experimental results and in-depth analyses indicate that our proposed method significantly enhances the performance of current matching models, such as the pretrained BERT and GPT-2, across two conversational corpora.

To summarize, our contributions are threefold:
First, we proposed a query-bag selection module, which leverages contrastive learning to rank and select relevant queries from the unlabeled corpus to form the query-bag with little noise.
Second, we proposed a query-bag fusion module to fuse mutual information of the queries and selected query-bags for query refinement.
Finally, we verified our proposed framework on various pretrained backbone models. 
Experimental results show that our proposed method can better use the refined query representation and improve existing models' performance in information-seeking conversations.

\section{Related Work}

\textbf{Retrieval-based Dialogue.}
Retrieval-based dialogue models have always been a challenging research area in natural language processing. In order to select the most relevant answers from the corpus, the previous method focus on learning to extract crucial matching information to calculate utterance relevance \cite{ zhou2018multi, gao2020learning, gao2021learning, chen2022multi,li2024multi}. 
DAM \cite{zhou2018multi} stacks multiple attention networks to model the interaction of multiple utterances and current responses.
BERT-FP \cite{han2021fine} utilizes large-scale pre-trained models (i.e., BERT) and leverages multi-task learning to learn the interactions of utterances and the utterance relevance classification to model discourse relevance and understands the semantic relevance and coherence between the dialogue context.
Other efforts focus on improving the quality of coarse-grained recalled candidates, such as utilizing a dense retrieval module and an interaction layer to obtain high-quality candidates, aiming to enhance the effectiveness of the final re-ranking stage ~\cite{lan2024exploring}.

\textbf{Information-seeking Conversation.}
The information-seeking conversation is a popular branch of retrieval-based dialogue. 
Most of the large-scale information-seeking systems in the industry are based on retrieval systems. 
Currently, the information-seeking conversation system highly relies on the quality and quantity of data. 
Therefore, it is necessary to pay attention to how to utilize more unlabeled data to boost the performance of the model (i.e., low-resource research). 
At the same time, transferring user questions to the model to achieve the goal of lifelong learning has become the trend in this research domain. 

For instance, \citeauthor{pan2021learning} enhances response candidates by using pseudo-relevance feedback (PRF) and proposes a reinforced selector to extract useful PRF terms, then rank the response with BERT for better performance. 
\citeauthor{gaur2022iseeq} propose an Information Seeking Question generator (ISEEQ) for generating Information Seeking Questions (ISQs) from short user queries with a knowledge graph to enrich the user query. 
IART \cite{yang2020iart} weights sentences by introducing users' intentions, which ultimately affects the ranking of answer choices. 
\citeauthor{yang2018response} incorporate external knowledge into deep neural models with pseudo-relevance feedback and QA correspondence knowledge distillation among information-seeking conversation systems. 
In terms of low-resource research, \citeauthor{voskarides2020query} classify  the user query into two categories whether it can be used to arrive at a better expression and modifies the query according to the classification results, to clarify user needs and reduce the annotation data required by the task. 
QRFA \cite{vakulenko2019qrfa} uses process mining technology to understand the structure of the user interaction process. 

Unlike the proposed methods, our framework targets to utilize the pseudo relevance feedback algorithm to select more equivalent queries with different expression to enhance the diversity of the training data and further fuse valuable information to boost the performance of the information-seeking conversational system.

\section{Methodology}\label{sec:method}
In this paper, we proposed a Query-bag Pseudo Relevance Feedback framework for information-seeking conversations.
Below we will first give an overview of the proposed framework and further illustrate the details of each module.
\input{latex/resource/fig1.tex}

\subsection{Overview}
We define the information-seeking conversation as a classification task. 
Given the user query ${Q}$ and the retrieval candidates ${B=\{b_1, b_2, ..., b_{k}}\}$, the classification model aims to rank the candidates based on the similarity between Q and B. 
The training objective is to minimize the cross-entropy loss: 
\begin{equation*}
\resizebox{0.48\textwidth}{!}{
    $\begin{aligned}
    \mathcal{L}_{CE} = \textstyle \sum_{(<Q, b_i>,y)}{-y\log(\hat{y})+(1-y)\log(1-\hat{y})}
    \end{aligned}$
}
\end{equation*}

We structure the QB-PRF framework as four sub-module: Representation Learning, Query-bag Selection (QBS), Query-bag Fusion (QBF), and Text Matching. We divide the model training into two stages. Stage 1 completes representation learning training, while Stage 2 completes the joint training of QBS, QBF, and text matching.
Here, we will provide a brief overview of our joint learning framework and further delve into more details.

\subsection{Representation Learning}
To provide supervision signals for QBS training, we leverage unsupervised pre-training methods to obtain differentiated sentence representation.
Specifically, we apply the classic Variational Auto-Encoder (VAE) \cite{kingma2013auto} network which consists of the bi-GRU layers:
\begin{equation*}
    \begin{aligned}
    \mathcal{L}_{VAE}=E[logp_\theta(x|z)]-\beta D_{KL}(q_\phi(z|x)||p(z))
    \end{aligned}
\end{equation*}

To enable the model to learn distinctive representation, we also integrated InfoNCE loss $L_q$ into the training.
This ensures the model adjusts the distribution of the questions and enables similar queries to occur around in the same embedding space:
\begin{equation*}
    \begin{aligned}
    \mathcal{L}_{q}=-log\frac{exp(q\cdot k_+/\tau)}{\sum_{i=0}^{k}{q\cdot k_i/\tau}}
    \end{aligned}
\end{equation*}
where $k_+$ represents the positive query related to q in a batch, ${\tau}$ is the hyper-parameter, and ${k+1}$ is the batch size.

The training of the representation learning network is formulated as: 
\begin{equation*}
    \begin{aligned}
\mathcal{L}_{Stage1}=\mathcal{L}_{\text{VAE}}+\alpha \mathcal{L}_{q}
    \end{aligned}
\end{equation*}
The entire training of representation learning is completed independently, as seen in Stage 1 of Figure \ref{fig:qb_structure}.

Due to the different languages in our benchmark data, we trained two language versions of the VAE model. For the English version, we utilized a dataset of 1 million questions from the Ubuntu corpus \cite{lowe2015ubuntu}, while for the Chinese version, we collected 1 million questions from Baidu Zhidao\footnote{https://zhidao.baidu.com/}, one of the largest question-and-answer websites in China. 
With the VAE model, we can eventually learn a better representation of the queries.
We then utilize such representation to implement query-bag selection.

\subsection{Query-bag Selection}
As mentioned, QBS module is utilized to select query-bag with little noise for the refinement of the corresponding query. Different from baseline query retrieval, QBS expands the query scope, traditional retrieval only searches questions from annotated Q-A pairs, while QBS retrieves from the entire accessible question pool. It can also include unreferenced user questions as retrieval targets, as this data often comprises a significant and diverse amount.
With the representation learning network learns a distinctive representation of the queries, the QBS module learns to select valuable queries to form the query-bag via contrastive learning.


We leverage the pre-trained VAE model to encode all questions and utilize the dense retrieval model to request similar questions.  
Precisely, We store the encoded questions in the Faiss index, and then similarly use VAE to encode the input query ${Q}$. Subsequently, we utilize this encoding to retrieve the top-k query-bag candidates ${bi}$ from Faiss, which serve as input for QBS.
Then we concatenate the representations to form a series of query-question (Q-Q) pairs. 
We project the Q-Q pairs into a BiLSTM layer with a residual network to get the shallow interaction between the query and question. 
The representation is subsequently fed into a two-layer linear classifier to compute the probability that determines its selection into the query-bags:
\begin{equation*}
    \begin{aligned}
    \text{SN}(b_i)=\text{Softmax}(W_g(\text{BiLSTM}(Q\oplus b_i)) + b_g)
    \end{aligned}
\end{equation*}
where ${W_g}$ and ${b_g}$ are the parameters of the classification module
\footnote{We apply the BiLSTM network with residual layer based on experimental observations, where the BiLSTM network performs better than directly feeding the representation to linear layers.}.

We apply the re-ranking process during training to ensure the QBS module learns to distinguish valuable queries from non-valuable ones.
Specifically, the re-ranking process is supervised by InfoNCE loss ${\mathcal{L}_b}$ and a selection reward loss ${\mathcal{L}_{reward}}$. 
The InfoNCE loss, denoted as ${\mathcal{L}_b}$, is a technique in contrastive learning designed to ensure that related questions within a query-bag are grouped together while also maximizing their distinction from dissimilar ones:
\begin{equation*}
    \begin{aligned}
    \mathcal{L}_{b}=-log\frac{exp(q\cdot b_+/\tau)}{\sum_{i=0}^{M}{q\cdot b_i/\tau}}
    \end{aligned}
\end{equation*}
where ${M=B_s \cdot (B_s-1)}$, ${B_s}$ is the bag size, ${b_i}$ refers to the samples in the query-bag, $b_+$ represents the pseudo positive candidates related to q.

The selection reward loss ${\mathcal{L}_{reward}}$ is applied to represent the average gain of positive questions: 
\begin{equation*}
    \begin{aligned}
    \mathcal{L}_{reward}=-log\frac{\sum_{i=1}^{B_s}{{sim(q,SN(b_i)^{+})}}+\tau_1}{{\sum_{i=1}^{B_s}{SN(b_i)^{+})}}+\tau_2}
    \end{aligned}
\end{equation*}
where ${\tau_1, \tau_1}$ are the hyper-parameters.

With such implementation, the selected questions are expected to have higher confidence under the determined size of query bags. 
The two losses are jointly trained to reach a convergence state in the training process. It ensures the distance between the query and each positive question in the query-bag is as close as possible and enables the sum of the average distance between the selected questions and the original user query is the largest.
Finally, with the QBS module, given the user's query, we could retrieve top-k similar questions and re-rank the question candidates to form the corresponding query-bag.
The selected query-bag and the corresponding query will then be passed to the query-bag fusion module.

\subsection{Query-bag Fusion}
We utilized a two-layer transformer network to fuse the mutual information of the query-bag and the original retrieval user query to implement information fusion. 
We average the expression of user query interacts with each sentence from query-bag through the QBF network to obtain the components of query-bag fused the necessary tokens to the original question. 

Specifically, we take the user query ${q}$ and query-bag ${bi}$ and apply a cross-attention mechanism to conduct interaction between ${q}$ and ${b_1}$, ${b_2}$, ..., ${b_k}$ to receive a updated query representation ($CA(q)$).
Then we use the self-attention mechanism to make further interactions ($SA(q)$). 
We project the updated query embedding $CA(q)$ into query q, key k, and value v through the transformer layer and outputs the embedding as the fused representation of query-bags:
\begin{equation*}
    \begin{aligned}
    CA(q)=\textstyle \sum_{i=0}^{k}{\text{Cross-Attn}(q, b_i, b_i)}
    \end{aligned}
\end{equation*}
\begin{equation*}
    \begin{aligned}
    SA(q)=\text{Self-Attn}(\text{q=k=v=}q\oplus CA(q))
    \end{aligned}
\end{equation*}

With the QBF module, we can receive a refined representation containing query and query-bag information.
The representation is then passed to the matching model.
Note that we will not change the composition of the downstream classification network but only update the network's input to achieve information transition, which maximizes the flexibility and controllability of the downstream classification model.

\subsection{The Matching Modules}
Our proposed method applies to any model that accepts two inputs for classification without changing the framework of downstream tasks. 
We achieve the maximization of the performance of the matching model through jointly training for query-bag selection, query-bag fusion, and matching model, as depicted in Stage 2 of Figure \ref{fig:qb_structure}:
\begin{equation*}
\resizebox{0.48\textwidth}{!}{%
    $\begin{aligned}
    \mathcal{L}_{Stage2} =\lambda_1\cdot \mathcal{L}_{b}+(1-\lambda_1)\cdot \mathcal{L}_{reward}+\lambda_2\cdot \mathcal{L}_{CE}
    \end{aligned}$
}
\end{equation*}
where ${\mathcal{L}_{CE}}$ is the cross entropy loss that supervises the matching procedure, ${\lambda_1}$, ${\lambda_2}$ are the hyper-parameters\footnote{We set ${\lambda_1}$, ${\lambda_2}$ to 0.5 and 0.1 respectively.}.

\section{Experiment}
\input{latex/resource/experiment_result}
\input{latex/resource/experiment_result_2}

\subsection{Dataset}
We explore the effect of QBS and QBF on two conversation corpora.
In order to facilitate the comparison with manually labeled query-bags, we verified the effectiveness of our proposed framework on English dataset, Quora.
Since Quora has pre-defined query-bag data, it is suitable for our experimental verification. 
Besides Quora, we also sought another Chinese dataset, LaiYe. 
The LaiYe dataset is collected from \url{https://laiye.com}.
The oracle query-bag statistics of LaiYe and Quora datasets are listed in Table \ref{tab:data_count}.
We introduce the details of the dataset below.

 The \textbf{Quora} dataset is derived from the question detection task. 
The query pair in the dataset identifies whether two user queries relate to the same answer. 
To construct the training and validation set, we arranged the questions into Query-Query pairs (Q-Q pairs) and randomly selected negative samples from other query-bags. 
The positive/negative ratio of the training set and the verification set is 1:1, and the positive/negative balance of the test set is 1:9.
 The \textbf{LaiYe} dialogue dataset is from the Laiye human-machine dialogue platform. 
We form the query-bag with user query and several similar questions corresponding to the same answer. 
Domain expert labels each query in the query-bag. 
We form user query and each question in query-bags into Q-Q pairs for training.

\subsection{Backbone Models}
To test our proposed framework, we select two classic pretrained models with superior performance on response selection tasks.
\textbf{BERT}, is the bidirectional encoder representation from transformers. 
In the pre-train era, verifying the impact of our QBS and QBF strategies on the pre-train model is necessary. 
We initialize the model with the BERT pre-defined parameters on Huggingface.
\textbf{GPT-2}, is a state-of-the-art language generation model that uses deep learning to generate human-like text. Its ability to produce coherent and contextually relevant responses has made it one of the most popular and widely-used generative models in natural language processing.

\subsection{Baselines}
To prove the effectiveness of our framework, we propose three baselines. The LLM Matching with large language models, the Q-Q matching without using any extra information, the ANCE-PRF matching that uses pseudo relevance feedback (PRF) to improve query representations for text matching.

\textbf{LLM Matching}. To observe the performance and limitations of large language models in text matching tasks, we conducted evaluations using OpenAI's GPT-3.5 interface. For each query in the test set paired with 10 questions, we instructed GPT-3.5 to provide relevance scores for these 10 pairs.

\textbf{Q-Q Matching}. The traditional models of text matching are based on Q-Q pairs. 
The BERT and GPT-2 use the training set for fine-tuning based on the pre-trained parameters. After the model training, the prepared Q-Q pairs will be fed into the matching model to obtain the matching probability of the two questions and compare them with the other baselines we proposed.

\textbf{ANCE-PRF Matching}. To verify the effectiveness of our proposal, we compare with a recent baseline ANCE-PRF \cite{yu2021improving} that automatically retrieves queries from a dense retrieval model, ANCE, and then uses a BERT encoder to consume the origin query and the top retrieved positive queries to produce a better query embedding, and then performs Q-Q matching. In the work of ANCE-PRF, all retrieved candidates are concatenated together into an extended text segment, which is then fed into BERT for a refined representation.

\subsection{Training and Evaluation}
Both models are implemented using PyTorch. 
For VAE implementations, the embedding is initialized the same way as BERT with the pre-train vector. 
The noise is sampled from ${\epsilon \sim Normal(0, 1)}$. The batch size is set to 128.
The encoder is a bidirectional GRU with 512 hidden units in each direction. 
The encoder and decoder units are both GRUs with 512 hidden units. 
The hidden layers of the encoder and decoder are set to 2 and 1, respectively. 
Before reconstruction, the ratio of the mask tokens is set to 1.0. 
For the QBS module, the hidden size of LSTM is set to 128, and the dropout is set to 0.5. 
For the transformer of the QBF module, the hidden size, attention heads, the number of hidden layers, and the dropout are set to 768, 8, 2, and 0.1, respectively. 
The max length of the sequence is 30 in LaiYe and 50 in Quora. 
For the hyperparameters in loss function, ${\tau}$, ${\tau_1}$, ${\tau_2}$ are set to 0.7, 0.0001, 0.001. 
For the training process, we then set the learning rate lr to 1e-5. We validate the matching model using the held-out validation set every 0.1 epoch.
More implementation details can be found in Section \ref{sec:method}.

For evaluation, we adopt standard metrics to assess retrieval performance. 
These metrics include Recall ($R@k$), which evaluates how many of the total relevant items are captured in the top-k recommendations. 
Mean Reciprocal Rank (MRR) refers to the average of the reciprocal ranks of the first relevant item in the list of recommendations.

\section{Results and Discussion}
\input{latex/resource/qbs_exp}
Table \ref{tab:experirment_result} 
demonstrates the main experimental results on the LaiYe, and Quora datasets on the baselines, QBS and QBF modules with different strategies applied. 
Table \ref{tab:experirment_result_2} refers to the performance variation with different candidates selection in our proposed framework.
Figure \ref{fig:QBS_exp_label}(a) and \ref{fig:QBS_exp_label}(b) shows the result of different dense retrieval settings.
Figure \ref{fig:QBS_exp_label}(c) reveals the accuracy of query selection in QBS module.
We analyze these results from the following perspectives.

\subsection{Comparison with Baselines}

Based on the results in Table \ref{tab:experirment_result}, we found that the PRF framework effectively increase the performance of matching models in the information-seeking system. 
Specifically, the matching models using query-bags are better than the simple BERT and GPT-2 models.
The improvement is more significant on the LaiYe dataset, while the increase on BERT is notable than the GPT-2 model. Possibly due to GPT's training limitation of only observing left-side context, its performance in tasks requiring global context comprehension, such as semantic understanding and text classification, may not be as superior as that of BERT.
During training, we also found that the model will converge to the best status faster and much more stable when using the PRF framework.
Despite being a powerful language model with broad capabilities, we observed that GPT-3.5 yields significantly poorer results compared to BERT. This discrepancy could be attributed to several factors, including BERT's pre-training objectives, token-level representation, fine-tuning mechanism, and model characteristics, all of which may contribute to its superior performance in classification tasks.


The results of ANCE-PRF slightly outperform BERT and GPT-2, indicating that ANCE has enhanced the representation of the original query to some extent. However, the results of ANCE-PRF are not as good as those of our model, suggesting that our model has improvements in both queries filtering and queries fusion. This can be observed through the results of ANCE-PRF without QBS. From the results of QB-PRF without QBS and BERT with ANCE-PRF, it is evident that the performance significantly decreases when QBS operation is not conducted. This indirectly demonstrates the effectiveness of the QBS module. Additionally, to validate the effectiveness of QBF under different strategies, we also performed sum and mean operations on the query-bags selected by QBS with the original query. The results indicate that these two operations introduce noticeable instability in the final performance. This experiment also confirms the robustness of our QBF module.

\subsection{Ablation Study}
We conduct two ablation studies to examine different components of our framework. 
Firstly, we omit the QBS and incorporate the top-k query candidates obtained from dense retrieval into fusion for query refinement. 
Secondly, we eliminate the QBF process, where the query-bag generated by the QBS module is fused with the original query using summation to create a refined representation. Subsequently, this refined query representation is directly fed into the matching models to enhance performance.
The ablation study results in Table~\ref{tab:experirment_result} indicate that each component of the QB-PRF framework plays a critical role in enhancing its overall effectiveness. 
The presence of the QBS module is crucial for precision optimization, as its omission results in a notable precision increase in query-bag and aids in more effectively enhancing the representation of the query.
It shows some improvements when we look into the results of fusion after retrieval without QBS.
That infers the effectiveness of VAE pre-train with contrastive learning on the large-scale dataset.
There is a significant drop in R10@1 when the QBF is excluded, underscoring its vital role in refining queries for subsequent text matching processes.

\subsection{Size of Query-bag}
According to Table \ref{tab:experirment_result}, Table \ref{tab:experirment_result_2}, Figure \ref{fig:QBS_exp_label}(a) and Figure \ref{fig:QBS_exp_label}(b), several observations are worth mentioning. 
In Table \ref{tab:experirment_result_2}, we evaluate our framework with different counts of retrieval candidates. 
As shown, recall of 3 or 5 also brings better results, and it shows that we can always benefit from similar questions even if the count is limited.
Experiments have verified query-bags' effectiveness, but the quantity and quality of query-bags do not always bring maximum improvement. 

According to Figure \ref{fig:QBS_exp_label}(a), we can see that the average query-bag size of LaiYe is relatively large.
For the BERT model, the metric ${R_{10}@1}$ lifts an increase of 6.1 percent.
Compared to LaiYe dataset, in Quora, with a smaller average query-bag size, the same metric lifts 4.5 percent on BERT.
When verifying strong models such as BERT, even when the size of query-bags is small, the performance on the Quora dataset is still strong. 
In combination with Table \ref{tab:experirment_result} and Figure \ref{fig:QBS_exp_label}(a), with the increase of question candidates, the size of the recalled query-bag gradually increases. 
Before the QBS process, there was more noise in the query-bag candidates, especially in the Quora dataset. The proportion of real query-bags in the recall of 10 candidates is only 9\%. 
Many noises directly lead to the decline of the fusion effect, and certain noises can be tolerated when the diversity of queries is substantial. 
In Figure \ref{fig:QBS_exp_label}(b), as the top-k value increases, the accuracy of the query candidates decreases, indicating a higher presence of noise within the query-bag. And our QBS module is designed specifically to reduce noise.
In practical applications, we need to select different top-k settings according to the dataset's characteristics, such as the LaiYe dataset having substantial diversity. 
The specified settings can achieve the best results.


\subsection{QBS Performance Tracking}

QBS, QBF, and the Matching module are jointly trained. In order to validate the effectiveness of QBS in selecting query-bags, we track the accuracy of the QBS module. For the top-k candidate queries obtained from dense retrieval, we provide ground truth annotations based on their semantic consistency with the query. During performance evaluation at several evaluation steps within each epoch, we simultaneously assess the accuracy of QBS in selecting query-bags. As indicated by the blue curve in Figure \ref{fig:QBS_exp_label}(c), the performance of QBS continues to improve over time.

\section{Conslusion}
In this paper, we propose a Query-bag Pseudo Relevance Feedback framework for an information-seeking conversation system.
To enable the framework to learn from diversified representation, we proposed a QBS module to learn to select relevant queries from question set to form the query-bag.
Such a module enriches the original query through additional information and enables the framework to learn distinctive representation.
With the selected query-bag, we then utilized the proposed QBF module to fuse the information of query-bags and eventually receive a distinctive query representation.
With the refined query representation, we can then perform the downstream text-matching task.
To assess the efficacy of our suggested framework, we conducted thorough experiments utilizing competitive pretrained backbone models. The experimental outcomes validate the effectiveness of our approach in enhancing text-matching performance across the pretrained models.

\section*{Limitations}
There are two main limitations in our approach.
First, the retrieved query-bags candidates are relatively scarce. 
It may cause the PRF framework to have imbalanced query-bag quantity and quality.
In the future, we shall verify the efficiency of our method after improving the recall of retrieval. 
Second, since, our framework applies to any neural model for any task since it proves the efficiency of information fusion with relevant data. 
It is valuable to manipulate in other tasks, such as retrieval-based dialogue or retrieval-augmented generation.

\bibliography{custom}

\end{document}

%% file: latex/resource/fig1.tex
\begin{figure*}[ht]
    \centering
    \includegraphics[width=\linewidth]
    {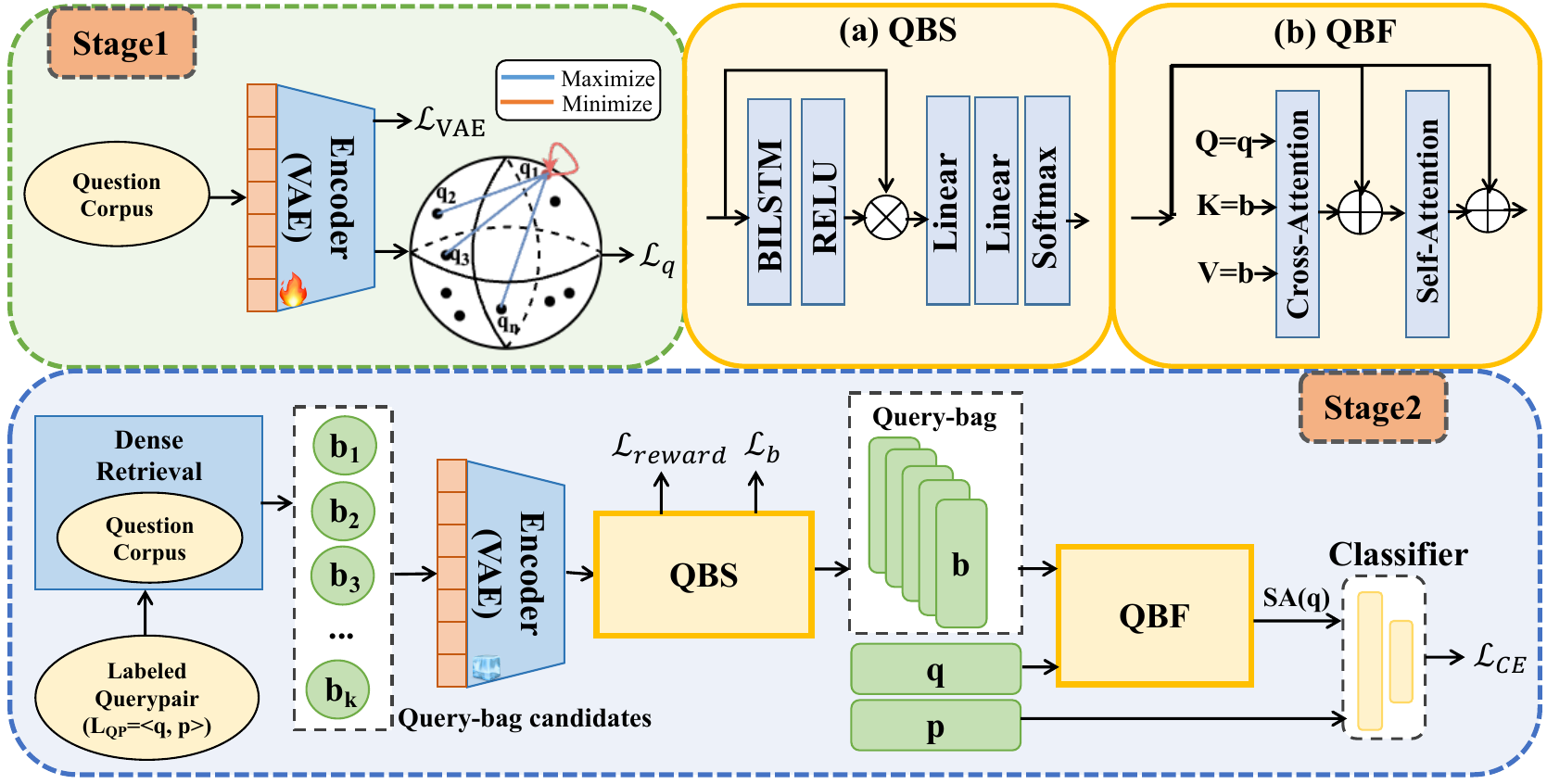}
    \caption{The structure of the QB-PRF framework. (a) QBS module, employs the $\mathcal{L}_{reward}$ and $\mathcal{L}_{b}$ to supervise the selection of similar queries. (b) QBF module, utilizes the multidimensional attention for a refined query representation.}
    \label{fig:qb_structure}
\end{figure*}

%% file: latex/resource/experiment_result.tex
\begin{table*}[t]
\centering
\small
\begin{tabular}{ccccccccc}
\toprule
\textbf{}          &  \textbf{Train} &  \textbf{Valid} &  \textbf{Test} & \textbf{Queries}        & \textbf{Query-bags} & \textbf{Average Query-bags} & \textbf{Max} & \textbf{Min}\\ \toprule
\textbf{LaiYe} & 425,310                            & 40,000                           & 40,000                           &  1,187,928 & 257,352      & 11.20  &171 &10                                                      \\
\textbf{Quora}     & 56,294                             & 5,536                            & 10,000                           & 45,903                         & 12,177        & 8.43    &48 &4                                                    \\ \bottomrule
\end{tabular}
\caption{The oracle query-bag statistic of LaiYe and Quora dataset. The train/valid/test is the Q-Q number. The max/min is the number of queries in the query-bags.}
 \label{tab:data_count}
\end{table*}

\begin{table*}[t!]
    \centering
    \resizebox{1\textwidth}{!}{
        \begin{tabular}{l ccccc | ccccc}
            \toprule
{\textbf{Models}} & \multicolumn{5}{c}{\textbf{Quora}} & \multicolumn{5}{c}{\textbf{LaiYe}} \\
\cmidrule(lr){2-6} \cmidrule(lr){7-11}
& \textbf{MRR} & \textbf{R10@1} & \textbf{R10@2} & \textbf{R10@5} & \textbf{R2@1} & \textbf{MRR} & \textbf{R10@1} & \textbf{R10@2} & \textbf{R10@5} & \textbf{R2@1} \\ 
\midrule
\textbf{GPT-3.5} &0.829    &0.467    &0.494     &0.563    &0.560    &0.692    &0.559    &0.724    &0.873    &0.745 \\ 
\midrule
\textbf{QB-PRF(BERT)} &\textbf{0.963}    &\textbf{0.915}    &\textbf{0.945}     &\textbf{0.983}    &\textbf{0.974}    &0.895    &\textbf{0.727}    &\textbf{0.866}    &\textbf{0.972}    &\textbf{0.850} \\ 
\textbf{Q-Q(BERT)} &0.927    &0.870    &0.918    &0.972    &0.958    &0.899      &0.666    &0.838    &0.955    &0.812 \\  
\textbf{ANCE-PRF(BERT)} &0.942    &0.880    &0.927    &0.976    &0.963     &0.897    &0.694    &0.848    &0.969    &0.848 \\  
\midrule
\multicolumn{4}{@{}l}{\emph{Ablation models}} \\  
\textbf{QB-PRF(BERT) w/o QBS} &0.920    &0.877    &0.920     &0.973    &0.963    &0.894    &0.670    &0.845    &0.960   &0.820 \\  
\textbf{QB-PRF(BERT) w/o QBF} &0.942    &0.866    &0.918     &0.978    &0.964    &0.839    &0.487    &0.668    &0.899    &0.784 \\  
            \toprule
        \textbf{QB-PRF(GPT-2)} &0.905    &\textbf{0.770}    &\textbf{0.841}     &\textbf{0.953}    &0.919    &\textbf{0.869}    &\textbf{0.673}    &\textbf{0.818}    &\textbf{0.957}    &\textbf{0.836} \\  
\textbf{Q-Q(GPT-2)} &0.895    &0.747    &0.838    &0.944    &0.919     &0.865    &0.658    &0.815    &0.956    &0.832 \\  
\textbf{ANCE-PRF(GPT-2)} &0.913    &0.753    &0.829    &0.940    &0.920     &0.862    &0.656    &0.802    &0.949    &0.825 \\ \bottomrule
        \end{tabular}}
    \caption{Automatic evaluation results on two datasets with different backbone models. \textbf{Bold} numbers indicate statistically \textit{significant} (p-value $< 0.05$) improvements over the second best result.}
    \label{tab:experirment_result}
\end{table*}

%% file: latex/resource/experiment_result_2.tex
\begin{table}[t]
 \centering
 \footnotesize
 \begin{tabular}{ccccccccccc}
   \toprule
   
   {\textbf{Top-k}}& \textbf{MRR} & \textbf{R10@1} & \textbf{R10@2} & \textbf{R10@5} & \textbf{R2@1} \\ \cmidrule{1-6}
   {{\textbf{1}}} 
   &0.874    &0.658    &0.833    &0.955    &0.814    \\
   {{\textbf{3}}} 
    &0.878      &0.659    &0.842    &0.959    &0.818\\ 
   {{\textbf{5}}} 
       &0.878   &0.669    &0.848    &0.960    &0.828\\ 

   {{\textbf{10}}} 
      &\textbf{0.895}   &\textbf{0.727}   &\textbf{0.866}   &\textbf{0.972}   &\textbf{0.850}\\ 
   \bottomrule

 \end{tabular}
 \caption{The performance of matching model with QB-PRF(BERT) when using different retrieval candidates, i.e., k . It can be seen that top 10 consistently achieves the best performance.}
 \label{tab:experirment_result_2}
\end{table}

%% file: latex/resource/qbs_exp.tex
\begin{figure*}[ht]
\centering
\includegraphics[width=0.9\linewidth]{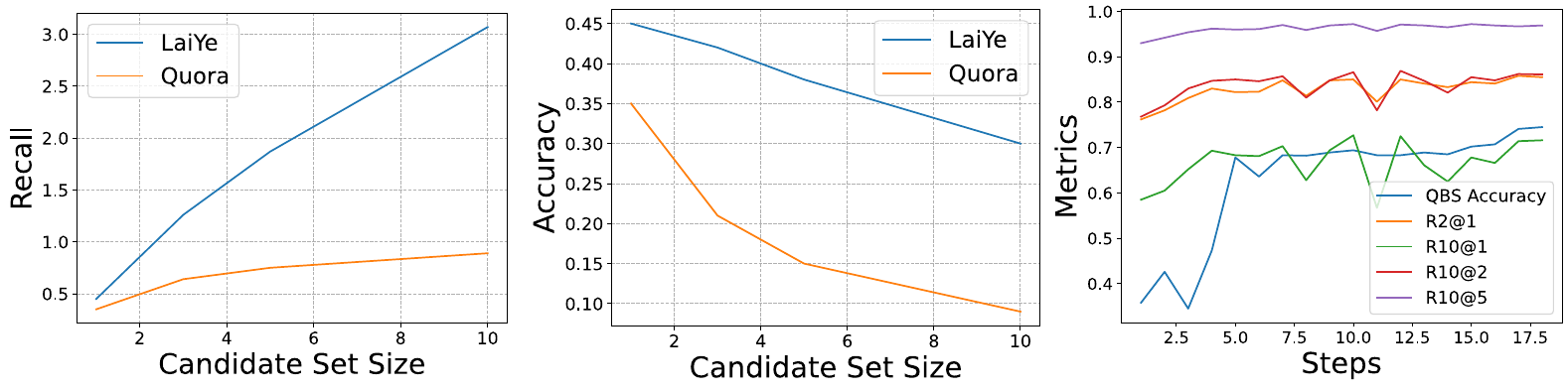}
    \caption{(a) Recall denotes the size of query-bag in datasets with regard to different candidate set size after dense retrieval in Figure \ref{fig:qb_structure}.
    (b) Accuracy refers to the proportion that belong to the query-bag with different candidate set size.
    (c) Performance on different evaluation steps. We evaluate the performance every 0.1 epoch.
    }
\label{fig:QBS_exp_label}
\end{figure*} 